# Chaotic dynamics of resting ventilatory flow in humans assessed through noise titration

Marc Wysocki (1,2)
Marie-Noëlle Fiamma (3)
Christian Straus (3,4)
Chi-Sang Poon (5)
Thomas Similowski (2, 3)

1. Hamilton Medical AG, Via Nova, CH-7403 Rhäzuns, Switzerland

2. Laboratoire de Physiopathologie Respiratoire, Service de Pneumologie et R_animation, Groupe Hospitalier Pitié-Salpêtrière, Assistance Publique-Hôpitaux de Paris

3. UPRES EA2397, Université Paris VI Pierre et Marie Curie, Paris, France.

4. Service Central d'Explorations Fonctionnelles Respiratoires, Groupe Hospitalier Pitié-Salpêtrière, Assistance Publique-Hôpitaux de Paris

5. Harvard-MIT Division of Health Sciences and Technology, Massachusetts Institute of Technology, Cambrige, MA, USA

**correspondence**
Thomas SIMILOWSKI, MD, PhD
Service de Pneumologie et Réanimation
Groupe Hospitalier Pitié-Salpêrière
47-83 Bd de l'Hôpital
75651 Paris Cedex 13
France
Tel: 33 1 42176797
Fax: 33 1 42176708
email : thomas.similowski@psl.ap-hop-paris.fr
**reprints**
Marc WYSOCKI, MD
Head of Medical Research
HAMILTON MEDICAL AG
Via Nova
CH-7403 Rhäzuns
Switzerland
Tel: +41 81 660 63 70
Fax: +41 81 660 60 20
email : mwysocki@hamilton-medical.ch

**Abstract**

The mammalian ventilatory behaviour exhibits nonlinear dynamics as reflected by certain nonlinearity or complexity indicators (e.g. correlation dimension, approximate entropy, Lyapunov exponents...) but this is not sufficient to determine its possible chaotic nature. To address this, we applied the noise titration technique, previously shown to discern and quantify chaos in short and noisy time series, to ventilatory flow recordings obtained in quietly breathing normal humans. Nine subjects (8 men and 1 woman, 24-42 yrs) were studied during 15-minute epochs of ventilatory steady-state (10.1±3.0 breaths/minute, tidal volume 0.63±0.2L). Noise titration applied to the unfiltered signals subsampled at 5 Hz detected nonlinearity in all cases (noise limit 20.2±12.5%). Noise limit values were weakly correlated to the correlation dimension and the largest Lyapunov exponent of the signals. This study shows that the noise titration approach evidences a chaotic dimension to the behavior of ventilatory flow over time in normal humans during tidal breathing.

# 1. Introduction

In mammals, the ventilatory behavior is not entirely periodic but may exhibit complexity as do many other biological functions (see review in Elbert et al., 1994). Ventilatory complexity may occur as a result of the interactions between several respiratory pattern generating neuronal networks (Ballanyi et al., 1999; Smith et al., 2000) or feedback modulations by mechanical and chemical afferents (Sammon, 1994b; Van den Aardweg and Karemaker, 2002) subject to corticospinal disruptions (Davenport and Reep, 1995). Such nonlinear interactions are conducive for complexity to occur, spontaneously or in response to external stimulation (Sammon, 1994b). Indeed, compound neural oscillator models of the mammalian respiratory rhythm can exhibit complex dynamical behaviors under various periodic inputs (Matsugu et al., 1998).

Previous studies have characterized ventilation with certain indicators of nonlinear determinism or complexity (see Donaldson, 1992; Hughson et al., 1995; Fortrat et al., 1997; Small et al., 1999; Dang-Vu et al., 2000; Burioka et al., 2002; Miyata et al., 2002; Yeragani et al., 2002; Akay et al., 2003; Akay and Sekine, 2004; Baldwin et al., 2004; Miyata et al., 2004). However, such descriptors are typically highly sensitive to noise, making them particularly misleading when applied to short and noisy biological time series (Barahona and Poon, 1996; Bruce, 1996). One strategy to infer the presence of nonlinear determinism in a noisy signal is by statistical comparison with randomly generated "surrogate" data (Theiler et al., 1992). Small et al. (1999) estimated correlation dimensions in infants using the surrogate data approach and described ventilation as a nonlinear system with a two-dimensional periodic orbit.

However, the presence of nonlinear determinism in a dynamical system is not synonymous with the presence of chaos (Poon and Barahona, 2001). Ascertaining the chaotic nature of respiratory complexity is important for the validation of respiratory system models such as those proposed by Dang-Vu et al. (2000), Small et al. (1999) and (Matsugu et al., 1998). The above limitations can be circumvented by the noise titration technique described by Poon and Barahona (2001) for the detection of chaotic dynamics in the presence of noise (see Methods and Discussion).

The aim of the present study was, therefore, to answer the question of the chaotic nature of respiratory complexity in this setting by applying the noise titration technique to ventilatory flow signals gathered in spontaneously breathing healthy humans.

## 2. Methods

### 2.1. *Subjects.*

Nine healthy subjects (8 men, 1 woman, age 24-42 years) were studied after legal and ethical clearance by the appropriate authorities, according to the French law. The subjects provided written consent to participate, were informed of the methods used but were not informed that the purpose of the study was to describe their ventilatory behaviour. They had been asked to refrain from consuming alcohol for 24 hours prior to the study, to refrain from using any psychotropic substance for 72 hours prior to the study, and to avoid major sleep deprivation. During the recording sessions, the subjects were comfortably seated in a lounge chair equipped with headrests. They were asked not to close their eyes, and not stare at a fixed point. They could listen to a musical piece that they were asked to choose as emotionnally and rhythmically neutral. No particular instruction regarding breathing was given.

### 2.2. *Procedures.*

Each experimental session lasted 30 minutes. The first 15 minutes were devoted to achieve a reasonably stable ventilatory state. The analysis of the ventilatory signal pertained only to the second block of 15 minutes (120 to 160 ventilatory cycles depending on the breathing frequency of the subjects). All the subjects were studied while breathing exclusively through the mouth, through a flanged rubber mouthpiece with a nose clip on. Because the respiratory route influences the pattern of breathing (Perez and Tobin, 1985), the measurements were repeated in a subset of three subjects breathing through a nasal mask with the mouth closed and then through a full facial mask.

### 2.3. *Measurements.*

Instantaneous ventilatory flow ($\dot{V}$) was recorded through a pneumotachograph (3700A series, Hans Rudolf, Kansas City, MO, USA or or VarFlex, Bicore, Irvine, CA, USA). End tidal $CO_2$ ($P_{ET}CO_2$) was measured from a side port of the mouthpiece (ML 205, AD Instruments, Castle Hill, Australia). The signals were not filtered. They were digitized at a 50 Hz sampling rate (AD Instruments, Castle Hill, Australia) and downloaded to a personal computer for subsequent analysis. For each recording, tidal volume ($V_T$), breathing frequency ($f_R$), inspiratory time ($T_I$) and total cycle time ($T_{TOT}$) were measured, and the frequency content of the signal was described using a 1024 points fast Fourier transform function (Hamming window, no overlap, zero values removed).

2.4. *Non-linear Analysis.*

*Noise titration.* The noise titration procedure was applied as described by Poon et al. (2001). Figure 1 provides a graphical explanation of the principle of the technique; a detailed mathematical description that can be found elsewhere (Barahona and Poon, 1996; Poon and Merrill, 1997; Poon and Barahona, 2001). In brief, this technique begins by an identification process capable of a robust and highly sensitive detection of deterministic dynamics (including chaos) within a given time series, even when it is noisy and of short duration (Barahona and Poon, 1996; Poon and Merrill, 1997). Specifically, several Volterra-Wiener-Korenberg series are generated, with different degrees of nonlinearity ($d$) and embedding dimensions Kappa ($K$), to produce a family of linear and nonlinear polynomial autoregressive models. The best linear model is obtained by adjusting the $K$ value with $d = 1$ to minimize the Akaike information–theoretic criterion. The best nonlinear model is obtained by sequentially increasing $K$ values with $d > 1$. Both models are then compared and the null hypothesis (linearity) is tested against the alternate hypothesis (non-linearity) using parametric (F-test) and nonparametric (Whitney–Mann) statistics (Barahona and Poon, 1996). If the null hypothesis is rejected with an alpha risk of 1%, namely if the experimental series is best described with a non-linear model, the titration process itself can be started. White noise of incrementally increasing standard deviation is added to the time-series until the null hypothesis cannot be rejected by the nonlinear detection algorithm described above (equivalence point). The standard deviation of the corresponding added noise is called "noise limit" (NL). A NL value above zero indicates the presence of chaos, and the value of NL provides an estimate of its intensity. A NL value equal to zero indicates that the series is not chaotic or that the chaos that it contained has already been neutralized by the background noise. This process can be compared to a chemical titration in which the concentration of an acid (chaos) in a solution (the time series) is determined from the quantity of added base (noise) that is necessary to neutralize the solution (Fig.1).

In the present study, the noise titration was applied to the ventilatory recordings using a specific routine developed under Matlab® V.6.5 R13 (MathWorks Inc, Natick, MA, USA). In order to find the strongest titration potential, a trial-and-error process was followed that included the testing of $K$ values of 4 to 6 and of nonlinear degrees of 3 to 5 (of note, the default settings of the routine used were a $K$ of 6 and a linear degree of 3). The 9 corresponding combinations were tested on each of the ventilatory flow recordings performed during mouth breathing (81 tests). They were also tested on each of the ventilatory flow recordings performed during exclusive nose breathing in the three subjects so studied, and during face mask breathing. This procedure was first performed on the signal as acquired, namely without filtering and at a 50 Hz sampling rate. Then, because oversampling can introduce co-linearities in the signal (Barahona and Poon, 1996), the data were subsampled at 5 Hz (Fig.2) and the full noise titration process was repeated to evaluate the impact of the sampling rate on the results.

*Other nonlinear descriptors* (see Appendix for more details). They were determined from the ventilatory flow data subsampled at 5 Hz (see Discussion for the justification of this choice), in two steps, using the Dataplore® software package (Datan, Teltow, Germany). The first step reconstructed the ventilatory flow attractor in phase-space. This was achieved, according to the embedding theorem formulated by Takens (Takens, 1980), by constituting a finished number of state vectors of $K$ constructed coordinates starting from points separated by a time delay $\tau$. The time delay was chosen as the first zero of the autocorrelation function (as already used in other studies of ventilation, e.g. Yeragani et al., 2002). The embedding dimension $K$ was determined according to Liebert et al. (1991) as the dimension allowing a phase-space reconstruction with a percentage of false nearest neighbours between 10 and 15%. The second step of this procedure consisted in the computation of numerical nonlinear descriptors. Firstly, the Lyapunov spectrum and the largest Lyapunov exponent (LLE, an index of the sensitivity of the system to its initial conditions) were calculated using the polynomial interpolation approach described by Briggs (1990). Secondly, the correlation dimension (CD, a fractal dimension reflecting the complexity of the attractor) was calculated from the correlation integral given by the Grassberger-Procaccia algorithm (Grassberger and Procaccia, 1983).

2.5. *Statistics.*

The two sets of noise limit values obtained at 50 Hz and 5 Hz were compared using the Student's paired t-test after verifying that their distributions were normal and homoscedastic (Prism®4.01, Graphpad Software, San Diego, CA, USA). Statistical associations between values were evaluated by calculating the Pearson product-moment correlation coefficient. The threshold for significance was set at a p value of 0.05.

### 3. Results

#### 3.1. *Ventilatory pattern.*

All the subjects exhibited a normal breathing pattern during mouth breathing, with a breathing frequency of 10.1 ± 3.0 breaths per minute ($T_{TOT}$ 6.51 ± 2.39 s), a $V_T$ of 0.63 ± 0.2 L, and a minute ventilation of 5.94 ± 0.81 L/min.. Inspiratory time ($T_I$) averaged 2.67 ± 0.73 s hence a $V_T/T_I$ ratio of 0.24 ± 0.03 L/s, and a $T_I/T_{TOT}$ ratio of 0.42 ± 0.08. The within-series coefficients of variation were 23.96 ± 9.07 % for ventilatory frequency, 19.36 ± 5.07 % for tidal volume, 18.76 ± 6.42 % for $T_I$. All the subjects had a normal $P_{ET}CO_2$ during the recordings (between 36 and 40 mmHg). During mouth breathing with a nose clip on, tidal volume was higher and breathing frequency slower than during nose breathing, as previously reported (Perez and Tobin, 1985), with values during face mask breathing similar to that measured during nose breathing.

#### 3.2. *Frequency content of the signal*

The mean power frequency of the ventilatory signal obtained with a 50 Hz sampling rate was 0.239 ± 0.072 Hz. The frequency at maximum was 0.199 ± 0.063 Hz. The maximal frequency ranged from 0.586 to 1.367 Hz (average 0.869 ± 0.233 Hz).

#### 3.3. *Noise titration.*

The effects of adding randomly generated noise of Gaussian distribution to the flow signal (Poon and Barahona, 2001) are depicted by Fig.3. During mouth breathing, the noise titration procedure applied to the ventilatory flow signal at its "native" acquisition frequency (50 Hz) yielded a noise limit above zero in 52 of the 81 computations performed (9 combinations of *K* and *d* in 9 subjects). The highest of these noise limit values obtained in each subject are listed in Table 1, together with the corresponding combination of parameters. In three subjects, no non-linearity was detected whatever the combination of parameter tested. Of note, a 40 Hz sampling rate did not influence these results (data not shown). After subsampling the signal at 5 Hz (Fig.2), the frequency of non-linearity detection was roughly similar (54 out of 81). However, a noise limit above zero could be found in the 9 subjects with an embedding dimension of 6 and a non-linear degree of 4 or 5. The average value of the highest noise limit value was significantly higher with the signal subsampled at 5 Hz than with the signal sampled at 50 Hz (20.2 ±12.5 % vs. 7.0 ± 3.8 %, respectively, p = 0.0001).

Positive noise titration values were also found during nose breathing and face mask breathing in the three subjects so studied (Table 2), using the same embedding dimension and nonlinear degree for the three breathing modalities in each of the subjects.

3.4. *Other nonlinear descriptors*

Applied to the ventilatory flow signal subsampled at 5 Hz, the computation process described in the methods section and in the appendix yielded numerical values compatible with the presence of chaos. This was true in all the subjects studied, under all the conditions tested, for both the largest Lyapunov exponent (0.2042 ± 0.1004, range 0.0225-0.3347 bits/iteration) and the correlation dimension (3.03 ± 0.43, range 2.12 - 3.43). Lyapunov spectra were homogeneous among the 9 subjects.

A significant linear correlation was found between the correlation dimension and the largest Lyapunov exponent (R = 0.539, 95%CI 0.013-0.831, p = 0.04). The values of the largest Lyapunov exponent were correlated with the values of the noise limit (R = 0.606, 95%CI 0.112-0.860, p = 0.01). This was also the case for the correlation dimension (R = 0.541, 95%CI 0.016-0.832, p = 0.04).

Of note, the distributions of the largest Lyapunov exponents and of the correlation dimensions were less scattered before subsampling than after. This did not translate in readily visible differences in phase portraits or return maps.

## 4. Discussion

This study shows that the noise titration approach evidences a chaotic dimension to the behavior of ventilatory flow over time in normal humans during tidal breathing. This finding should help understanding already available results, and should open new perspectives for the utilisation of nonlinear approaches to describe respiratory physiology in health and disease.

4.1. *Chaos detection.* As mentioned in the introduction, several approaches have been used to study the nonlinear dynamics of respiration. Donaldson (1992), in 8 adults during resting breathing, found that minute ventilation, $T_{TOT}$, expiratory time $T_E$, or $P_{ET}CO_2$ could be characterized by positive Lyapunov exponents ranging from 0.06 for $P_{ET}CO_2$ to 0.23 for $T_E$. Yeragani et al. (2002) reported higher "respiratory" approximate entropies and Lyapunov exponents in patients with panic disorders as compared to controls. The surrogate data approach was used by the team of Burioka et al. (2002; 2003) to calculate correlation dimensions and approximate entropy from respiratory movements recorded during sleep in healthy subjects. They observed a reduction in respiratory complexity in normal subjects breathing $CO_2$-enriched gas mixtures (Suyama et al., 2003) and in patients suffering from the obstructive sleep apnea syndrome (Miyata et al., 2002). In rats, Sammon et al. (1991; 1994b; 1994a) showed that "spiral" attractors and "horseshoe" return maps and fractal correlation dimensions characterised the ventilatory behaviour but disappeared in the absence of feedback loops namely after vagotomy. The use of approximate entropy also allowed Akay et al. (2003) to describe age-related changes in the complexity of respiratory pattern, and to demonstrate that an hypoxic insult to the brain resulted in a decreased respiratory complexity (Akay and Sekine, 2004). Fractal fluctuations have been described in spontaneously breathing healthy adult humans (Fadel et al., 2004) and a decrease in respiratory complexity with age has also been shown by Peng et al. (2002) through the analysis of fractal scaling exponents. Baldwin et al. (2004), exploring the role of sigh in respiratory control with a combination of linear and nonlinear approaches (including long-range memory studied from detrended fluctuation analysis) showed that sighs were followed by a "restoration" of respiratory complexity under the form of an increase in the range of points located within a defined attractor. Finally, complexity has also been observed from in vitro preparations isolating the mammalian inspiratory neural network (Del Negro et al., 2002), where increasing neuronal excitability modifies the pattern of oscillations in the neural output (from mixed-mode oscillations, quasiperiodicity and ultimately aperiodic, chaos-like activity).

These data did not prove that there is chaos in the mammalian ventilatory behaviour. Firstly, nonlinear characterization methods, including Lyapunov exponents and fractal analysis, are sensitive to noise and produce statistically biased estimates of signal properties when they are applied to short time series (Bruce, 1996). The standard chaos detection methods in fact have virtually zero noise tolerance (see discussion in Poon and Barahona, 2001). Secondly, any record of respiratory activity may include variability of several types : random uncorrelated, random correlated, periodic, and nonlinear deterministic. The first two are always present and create obstacles to a reliable mathematical detection of the latter two (Bruce, 1996). Thirdly, these methods, including the surrogate data approach, infer the presence of nonlinearity and of determinism in a signal, yet these conditions are necessary but not sufficient to ascertain chaos. With noise titration, the use of the Volterra-Wiener algorithm provides a sensitive indicator of nonlinearity, and the very nature of the process (controlled addition of noise to the data) provides the technique with immunity to noise contained in the observation. This algorithm has been tested and validated to be not only necessary, but also sufficient, to detect chaos in short times series derived from models depicting any of the routes to chaos (Poon and Barahona, 2001). These routes include the succession of quasiperiodic tori, that is likely the most appropriate to consider about the human ventilatory behaviour. Indeed, this phenomenon can explain the onset of chaotic fluctuations in coupled neural oscillators (Matsugu et al., 1998) and probably also in neural networks. Of note, this model provides, after the onset of chaos, phase portraits and Poincaré sections that are closely matched by those obtained from ventilatory signals (see Dang-Vu et al., 2000).

That we were able to detect chaos within resting ventilatory flow signals with noise titration is thus reassuring, and consolidates the notion that human ventilation, at rest and during wakefulness, is actually chaotic in nature. Of note in this regard, we found largest Lyapunov exponents similar to that found by Donaldson (1992) and correlation dimensions similar to that found by Suyama et al. (2003).

4.2. *Source of ventilatory chaos.* Identifying respiratory chaos from a flow signal thoraco-abdominal displacements (as in Small et al., 1999) tells nothing about its source. A neural source for respiratory chaos is a likely hypothesis, because it is widely accepted that the central respiratory command depends on at least two coupled oscillators (see review in Feldman et al., 2003; Mellen et al., 2003; see also Vasilakos et al., 2005). Compound neural oscillator models of the mammalian respiratory rhythm can exhibit complex dynamical behaviors under various periodic inputs (Matsugu et al., 1998), and the study of breath-to-breath variations in tidal volume, end-tidal $O_2$ and end-tidal $CO_2$ in infants by Cernelc et al. (2002) has evidenced long-range correlation properties consistent with the neural network model of the respiratory central pattern generator. Nonlinear activities exist in *in vitro* isolated preparations of the network that generates respiratory rhythm in mammals (Del Negro et al., 2002).

However, nonlinearities in the flow pattern could also be created, or altered, by the nature of the media to which the respiratory neural command is applied. In this view, short-term fluctuations of the impedance of the respiratory system have been described (Que et al., 2001). In this view also, the propagation of a front within an anisotropic media can produce chaos (Bar et al., 2000).

4.3. *Chaos quantification*. The relevance of identifying chaos within a biological go through depends on the possibility to quantify it, in order to compare sequential or categorical situations. Applied to biological signals (and particularly to short and noisy time series, as opposed to benchmark model systems), none of the available methods have been shown to meet the criteria expounded by Bruce (1996), namely give a small or null value to constant or periodic signals, to give the largest possible value to a stochastic signal, and, most importantly, to provide intermediate but increasing values to deterministic signals of increasing dimension. Noise titration does so and therefore seems ideally suited to study respiratory signals (applied to benchmark model systems, it faithfully follows the evolution of Lyapunov exponent with increasing chaoticity (Poon and Barahona, 2001)) and is immune to noise. However, the absolute noise limit depends on the characteristics of the signal and particularly on the noise floor. A change in noise limit can therefore stem from a change in chaoticity or in the noise floor. This cannot be readily ascertained, and it is thus of the utmost importance, in experiments intended to be analysed with noise titration, to maximise effort to control for experimental noise and to minimise variations in physiological noise through the achievement of steady state. Our study pertained to tidal breathing at rest only, in an homogenous set of subjects. It is reasonable to assume that there was no major intersubject variations in the noise floor. The distributions of the Lyapunov spectrums, the largest Lyapunov exponents, and the correlation dimensions among the subjects were relatively narrow. By contrast, we observed noise limit values stretching from 6 to 43% (Table 1). This thus indicates that different subjects may exhibit different level of respiratory chaos and that this is not well described by other nonlinear descriptors. This probably explains why we found only weak correlations between noise limit and the largest Lyapunov exponent or correlation dimension.

4.4 *Technical issues*. We studied the ventilatory behaviour of our subjects through the values of a non-filtered and non-integrated instantanous flow signal. This is important because filtering (for example see Yeragani et al., 2002, who found largest Lyapunov exponents smaller than ours -0.086 ± 0.018) and integration (to study tidal volume, for example in Donaldson, 1992) imply information loss and can distort the characterization of nonlinearities. A raw flow signal also gives the possibility to directly obtain the equi-spaced values necessary for nonlinear analysis, without corrections or assumptions. It must also be emphasized that the parameters used to reconstruct the attractor exert an important influence on the values of "traditional" nonlinear descriptors. This is the case for $\tau$, that we defined as the delay needed to reach the first zero of an autocorrelation function (Liebert and Schuster, 1989) rather than arbitrarily (Dang-Vu et al., 2000). These constraints are alleviated by the noise titration approach. Our study also illustrates the importance of sampling frequency on the results of chaos detection (Table 1, Fig. 2). Oversampling can introduce linearities in the data (Barahona and Poon, 1996). With a sampling frequency of 50 Hz, we failed to detect nonlinearities in several cases and the average noise limit values were significantly smaller than with a 5 Hz sampling rate. The top trace in Fig. 2 exhibits signs of oversampling (e.g. small spikes erratically distributed over the signal). In contrast, undersampling carries the risk of aliasing, and if the sampling frequency is less than the Nyquist frequency high frequency folding to low frequency can show up as nonlinearity. The frequency content of the signal must thus be checked before applying noise titration or other nonlinear analysis. In the present set of data, the maximal frequency present in the ventilatory signal was 1.367 Hz (Results, 3.2). Our 5 Hz subsampling frequency is therefore not a likely source of artificial nonlinearities. From a practical point of view, it seems reasonable to recommend the application of noise titration to respiratory signals sampled with a frequency above the Nyquist frequency but of the same order of magnitude. Finally, instrumentation changes the nature of the measured object. Thus, going from "natural respiration" (nose breathing) to "measured respiration" (mouthpiece and nose clip) changes breathing pattern (Perez and Tobin, 1985). Although we have too few data to conclude firmly on this point, our results suggest that the detection of respiratory chaos by noise titration may not be excessively sensitive to instrumentation and breathing route (Table 2).

4.5. *Conclusions and Perspectives*. The detection of "respiratory chaos" by the noise titration technique within short epochs of biological signals collected under the duress of human physiology experiments makes plausible the clinical applicability of this mathematical approach. Its quantitative nature also opens the possibility to use it as a descriptor of the effects of interventions (or disease) on respiratory control, in the perspective of the relationship between illness and the altered variability of complex systems (Macklem, 2002; Seely and Macklem, 2004) (see also Poon and Merrill, 1997 for an application of chaotic analysis to the characterisation of heart failure).

**Acknowledgements**

This study was funded in part by a "*Contrat de recherche triennal "Legs Poix" de la Chancellerie de l'Université de Paris*", by *Association pour le Développement et l'Organisation de la Recherche En Pneumologie* (ADOREP), Paris, France and by the "*Ministère de la Jeunesse, de l'Education Nationale et de la Recherche*", Paris, France (*Fonds national de la science, Action Concertée Incitative "Technologies pour la Santé", projet MOSAIQUE*).

Marie-Noëlle Fiamma was supported by a scholarship from the "*Ministère de la Jeunesse, de l'Education Nationale et de la Recherche*", Paris, France. Chi-Sang Poon was supported by U.S. National Institutes of Health grant HL075014.

We are grateful to Sylvain Thibault, Gila Benchetrit and Pierre Baconnier from the PRETA-TIMC-IMAG laboratory of the Université Joseph Fourier, Grenoble, France, for the help with nonlinear analysis that they kindly provided to Marie-Noelle Fiamma, and to Mr Zhi-de Deng from the Department of Electrical Engineering and Computer Science at M.I.T. for his help with noise titration computations.

**Table 1.** Highest noise limit values found in the 9 subjects during exclusive mouth breathing, at the two sampling frequency tested.

| Sampling Rate -> | 5 Hz | | | 50 Hz | | |
|---|---|---|---|---|---|---|
| **Subjects** | NL (%) | *K* | d | NL (%) | *K* | d |
| 1 | 9 | 6 | 5 | \ | 4,5,6 | 3,4,5 |
| 2 | 19 | 6 | 4,5 | 4 | 6 | 3 |
| 3 | 9 | 6 | 4 | 2 | 6 | 4 |
| 4 | 43 | 6 | 4 | 10 | 5,6 | 4,5 |
| 5 | 18 | 6 | 4 | 11 | 6 | 5 |
| 6 | 28 | 6 | 4 | 5 | 6 | 5 |
| 7 | 16 | 6 | 5 | \ | 4,5,6 | 3,4,5 |
| 8 | 6 | 6 | 4 | \ | 4,5,6 | 3,4,5 |
| 9 | 36 | 6 | 4 | 10 | 6 | 5 |
| mean | **20.4** | | | **7.0** | | |
| sd | **12.8** | | | **3.8** | | |

NL, noise limit (the symbol "**Error!**");
*K* embedding dimension providing the highest NL;
d, nonlinear degree providing the highest NL value.
When two values are provided for *K* and d, the NL was not affected by the corresponding parameter change.
The 50 Hz values were significantly lower than the 5 Hz values.

**Table 2.** Effects of the breathing route on the highest noise limit value and on the other nonlinear descriptors in the three subjects studied with three different breathing modalities (data calculated from the data subsampled at 5 Hz, see text for details).

|   | Highest Noise Limit (%) | | | Largest Lyapunov Exponent (bit/iteration) | | | Correlation Dimension | | |
|---|---|---|---|---|---|---|---|---|---|
|   | *M* | *N* | *mask* | *M* | *N* | *mask* | *M* | *N* | *mask* |
| 1 | 9 | 8 | 2 | 0.0225 | 0.1371 | 0.093 | 2.12 | 2.69 | 2.68 |
| 2 | 19 | 21 | 21 | 0.1634 | 0.2667 | 0.2672 | 2.59 | 2.51 | 2.93 |
| 3 | 19 | na | 20 | 0.2215 | na | 0.1308 | 2.93 | na | 2.94 |

M, mouth breathing; N, nose breathing (data not available -na- in one subject for technical reasons); mask, face mask breathing.

**Legends to the figures**

**Fig.1.** Schematic representation of the noise titration process (from Poon and Barahona, 2001)

**Fig.2.** Example, in one subject, of the effects of the subsampling procedure on the ventilatory flow signal.

The top tracing (A) represents the signal as sampled duirng the experimental acquisition (no filtering, A/D rate of 50 Hz). The bottom tracing (B) represents the same epoch after subsampling at 5 Hz.

**Fig.3.** Effects of the addition of randomly generated noise of Gaussian distribution on the ventilatory flow signal and on the noise titration process. The left column depicts the flow signal in three different conditions: (A) after the 5 Hz subsampling (same example as in Fig. 2) but before the addition of any noise; (B) after the addition of 20% of noise to the signal; (C) after the addition of 36% of noise to the point of noise titration where the test for nonlinear detection fails (see Fig.1), namely the statistical comparison of the linear and the nonlinear models does not reject the hypothesis of their identity anymore. Beside each flow tracing are the corresponding plots of C(r), the Akaike information-theoretic criterion or cost function, against the number r of polynomial terms of the linear (black) or nonlinear (grey) estimates of the time series using the Volterra-Wiener method (Barahona and Poon, 1996)(see Methods, 2.4).

**Appendix**

This appendix describes in more details the process followed to obtain the "traditional" non-linear descriptors of the ventilatory flow signal in this study. All the operations were performed on the signal after subsampling at 5 Hz, using the Dataplore® v 2.0.9 software (Datan, Teltow, Germany).

*First step : attractor reconstruction.* The technique of attractor reconstruction was based on the theorem formulated by Takens (Takens, 1980) allowing the preservation of all the properties of the original attractor. A series of state vectors of "m" coordinates starting from time values separated by a delay "τ" was constituted according to *$S_i$= ($x_i$+t, $x_i$+2.t, É.$x_i$+(m-1).t)*. The analysis pertaining to a finite and noisy data set rather than to an infinite noise-free data set, it was not possible to choose the time delay arbitrarily. Indeed, under such conditions, the time delay influences the quality of the reconstructed trajectory and thereby the values of the characteristic quantities under consideration (Liebert and Schuster, 1989). Among the many available possibilities to determine an appropriate time delay, we chose to use the time to the first zero of an autocorrelation function as our time delay (Liebert and Schuster, 1989). This approach is interesting because the points beyond this first zero point are independent. Of note, this method has already been used in the study of the non-linear behaviour of ventilation (Yeragani et al., 2004). In our study, the time delay for attractor reconstruction was determined separately for each recording rather than globally.

The proper reconstruction of an attractor depends on a dimension of phase-space sufficient to unfold the attractor (embedding dimension). We used the false nearest neighbors method (Kennel Matthew B et al., 1992) to determine the embedding dimension most appropriate to our data. In brief, two points in a space are sais "neighbors" if they are distant of less than "r", "r" being an arbitrary distance that is in fact equivalent to the time interval between two data points. If two points that are neighbors in a space of dimension m are not any more so in a space of dimension higher than m, they are called false neighbors. The false nearest neighbors method consists in eliminating the false neighbors through iterations performed with unitary stepwise increases of the embedding dimension "m". For each value of "m" a percentage of false nearest neighbors is determined, that decreases when m increases. We retained the "m" value producing a percentage of false nearest neighbors between 10 and 15%, a value considered the smallest possible while compatible with the calculation of the Lyapunov exponents (Liebert et al., 1991).

*Second step : numerical characterization of the reconstructed attractors.* Among the main characteristics of a chaotic system are its complexity and its sensitivity to the initial conditions

The correlation dimension is a fractal dimension that gives a measure of the complexity of the strange attractor of the studied system. To obtain the correlation dimension of our time series, we first computed the correlation integral C(r), using the Grassberger-Procaccia algorithm (Grassberger and Procaccia, 1983) implemented in the Dataplore¨ software. C(r) is equal to the number of neighbours for a distance "r" in space phase divided by the total number of points in the space raised at the power 2. To carry out the calculation of the number of neighbors for a given point, a cube of r dimension is centered on this point and the number of points contained in the cube is deducted. Iterations are then carried out for each subsequent point, the time delay τ (see first step, above) being used as the pitch value, and the total of neighbors being the sum of the neighbors found by each iteration. Within an interval delimited by fractions of "r", the logarithm of C(r) is proportional to the logarithm of r, and the slope of the relationship is the correlation dimension. In our study, the log C - log r relationship was determined for embedding dimension from 2 to 10. Nine curves were obtained. The "r" boundaries within which these curves were linear were visually determined and the corresponding correlation dimensions calculated. Correlation dimensions were graphed as a function of embedding dimension. This provided a bell-shaped curve, of which the top ordinate was taken as the correlation dimension of the considered attractor.

The sensitivity of the system to the initial conditions is described by the Lyapunov exponents, which quantify the mean rate of exponential divergence of nearby trajectories along various directions in phase-space. An m-dimensional dynamical system has m-Lyapunov exponents. Three conditions are necessary (but not sufficient) for the behaviour of a dynamical system to be compatible with the presence of chaos (Wolf et al., 1985; Rosenstein et al., 1993; Hongre et al., 1999). At least one Lyapunov exponent must be positive to explain the divergence of the trajectories, at least one must be negative to justify the folding up of the trajectories, and the sum of all the exponents must be negative to account for the dissipative nature of the system. If a system is chaotic, the value of the largest Lyapunov exponent is often considered as a means to quantify the degree of chaos in this system (Wolf et al., 1985). In this study, we used the improved form of Wolf's algorith described by Briggs (Briggs, 1990) and that relies on polynomial used for least-squares fitting. This approach requires the determination of three parameters: the embedding dimension, the number of neighbors, the degree of the polynomial of interpolation. The embedding dimension was taken from step 1 as described above. To determine the other two parameters, we tested 200 combinations of the number of neighbors (from 1 to 50) and of the polynomial degree (from 1 to 4) on three of our recordings (separate subjects). We selected the combinations providing the largest Lyapunov exponent in the 0-1 range (indeed, values above one does not seem compatible with the description of the ventilatory behaviour in humans, see Donaldson, 1992; Dang-Vu et al., 2000; Yeragani et al., 2002) and a Kaplan-Yorke dimension above 0 (Briggs, 1990; Dang-Vu and Delcarte, 2000). Finally, we retained for application to the other recordings the one combination providing the Kaplan-Yorke dimension that was nearest to the correlation dimension (see above). All in all, the largest Lyapunov exponents in our time series were calculated with a number of neighbours of thirty, a polynomial degree of two, and an embedding dimension specific to each individual series.

**Figure 1R1**
**Click here to download high resolution image**

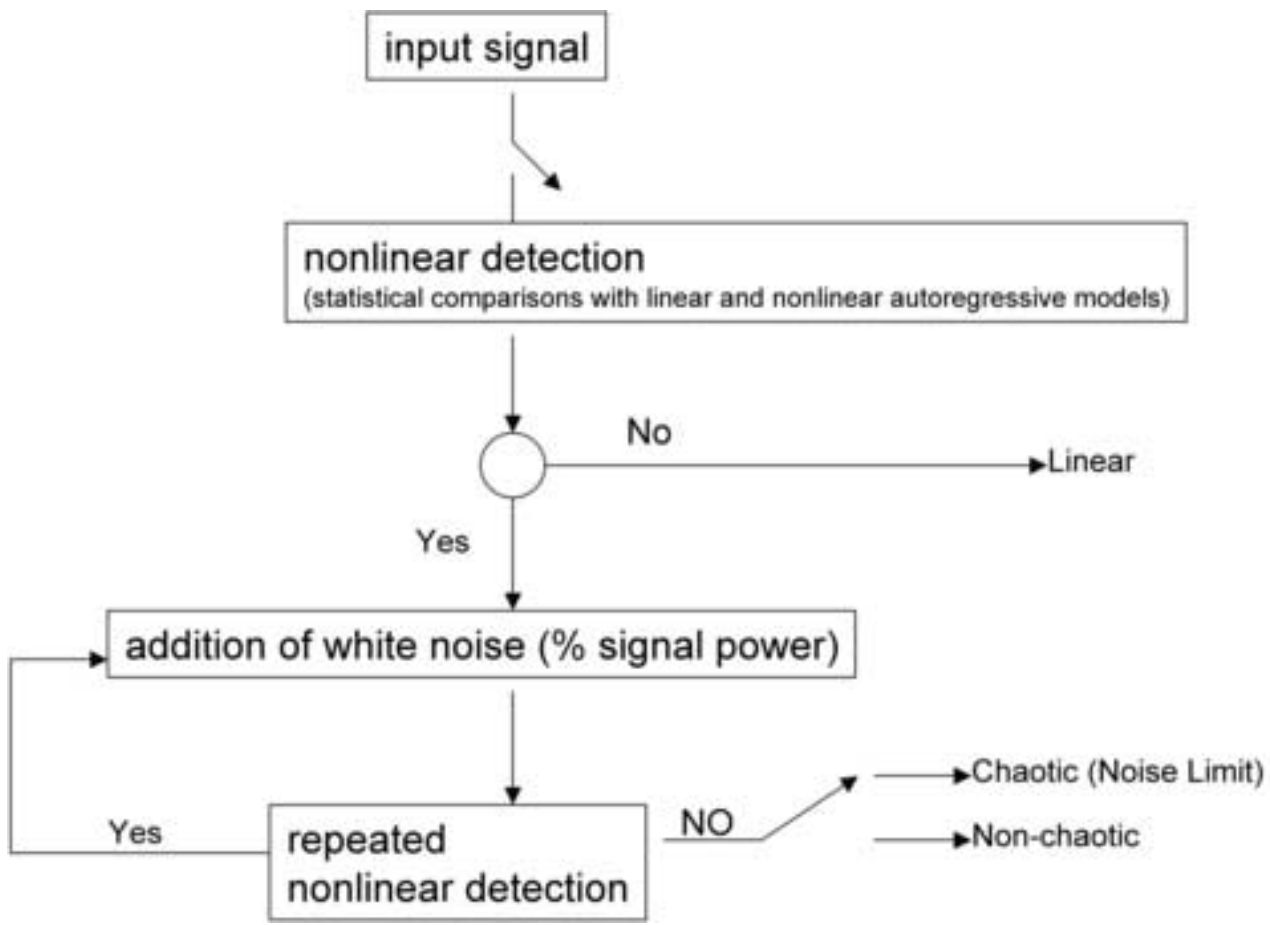

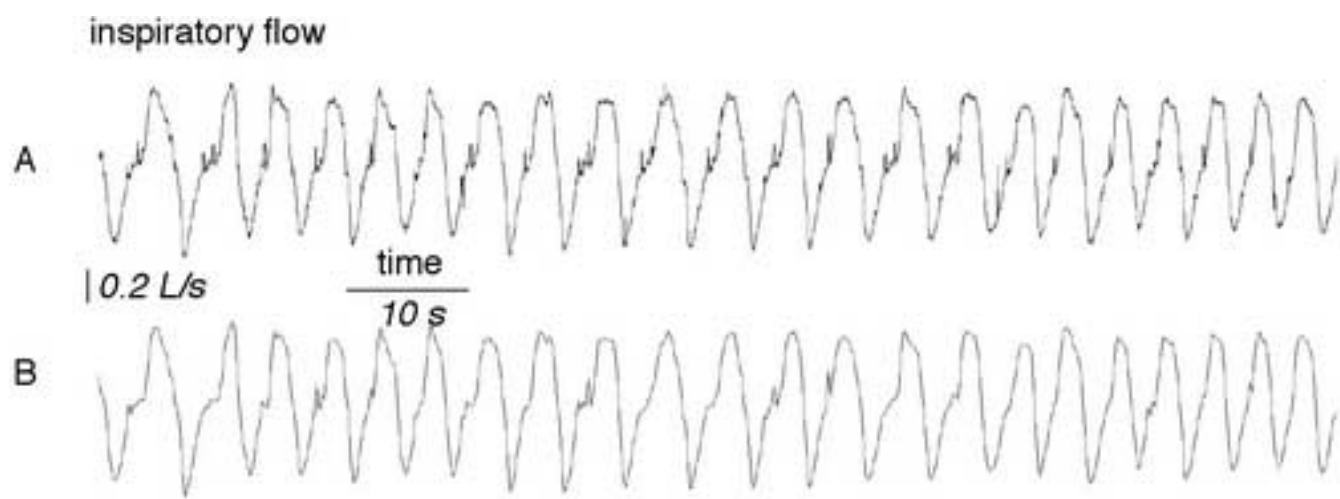

**Figure 3R1**
**Click here to download high resolution image**

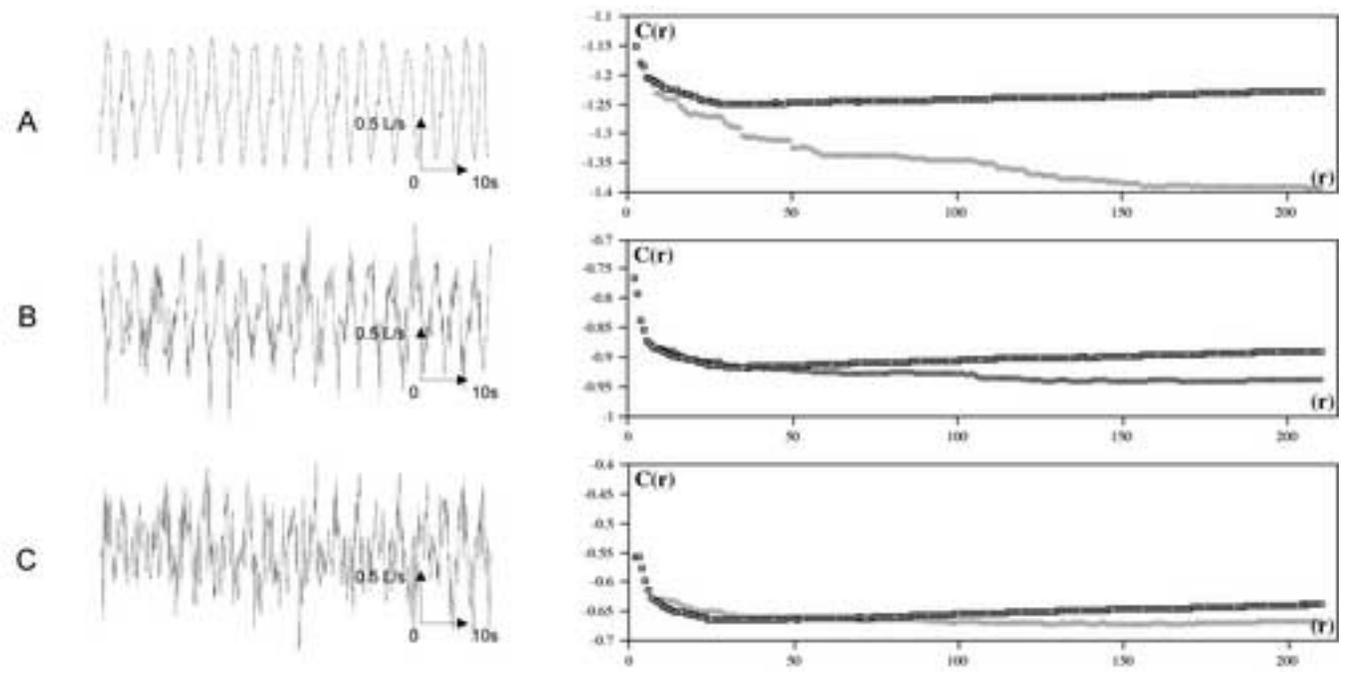